\documentclass[conference]{IEEEtran}
\IEEEoverridecommandlockouts
\usepackage{cite}
\usepackage{amsmath,amssymb,amsfonts}
\usepackage{graphicx}
\usepackage{textcomp}
\usepackage{xcolor}
\usepackage{url}

\begin{document}

\title{Mitigating the Grid Impact of Gigawatt-Scale AI Data Centers via High-Fidelity Model Flexibility}


\author{\IEEEauthorblockN{Eihab E. E. Ahmed}
\IEEEauthorblockA{\textit{Electrical and Computer Engineering} \\
\textit{San Diego State University and University of California San Diego}\\
San Diego, CA, USA \\
eelgailyelmuks5457@sdsu.edu}
\and
\IEEEauthorblockN{Saeed Manshadi}
\IEEEauthorblockA{\textit{Electrical and Computer Engineering} \\
\textit{San Diego State University}\\
San Diego, CA, USA \\
smanshadi@sdsu.edu}}

\maketitle
 
\begin{abstract}
Gigawatt-scale artificial intelligence data centers (AIDCs) are being connected at a single point of interconnection (POI) faster than transmission can be reinforced, yet studies of their flexibility rely on DC power flow and simplified load envelopes that omit voltage, cooling dynamics, and workload deadlines. This paper addresses this gap by embedding a high-fidelity AIDC model, with separate inference and training pools, hall thermal mass, PV, a battery energy storage system (BESS) with reactive capability, and a STATCOM, in a five-minute multi-period AC optimal power flow. Simulations on the RTS-GMLC test system show that active-power flexibility mitigates local price spikes and relieves congestion, whereas reactive support from the BESS inverter or a STATCOM is needed to keep the POI voltage away from sagging around its 0.95~pu limit. Sensitivity analyses show that cost-driven charging of large batteries can induce new morning congestion and that storage largely substitutes for wider workload deferral.


\end{abstract}

\begin{IEEEkeywords}
AI data centers, AC optimal power flow, large loads, flexibility, transmission congestion
\end{IEEEkeywords}

\section{Introduction}
U.S. data centers consumed 176~TWh in 2023, or 4.4\% of national electricity use, and artificial intelligence (AI) is expected to lift this figure to 325--580~TWh, or 6.7--12\%, by 2028~\cite{shehabi2024lbnl}. This growth will happen in large blocks. A single AIDC can place hundreds of megawatts to gigawatts on one bus and become site-ready within one to two years, while the transmission reinforcements it may require take much longer~\cite{kim2026siting}. NERC recommends that planners specify power-factor ranges and ramp limits for large loads~\cite{nerc2026rglargeloads}.

 Training jobs often tolerate deferral, the thermal mass of the data hall allows cooling to be rescheduled, and on-site batteries and photovoltaics (PV) can reshape the draw at the POI. Several studies have valued this flexibility on transmission networks. Spatial migration and temporal shifting of workload relieve congestion and renewable curtailment in DC optimal power flow (OPF) and security-constrained unit commitment~\cite{wan2026spatial,wan2026scuc}, and battery-assisted operation keeps an AIDC within its interconnection limits~\cite{lu2026battery}. Planning studies use the same approach to widen the feasible siting options~\cite{kim2026siting}. These studies represent the transmission network with DC power flow or omit it, so voltage magnitude and reactive power do not enter the analysis. Most of them also reduce the facility to a flexible share of its load or a standardized operating envelope, although cooling dynamics and service-level constraints may limit the flexibility a facility can actually deliver. Reactive support from battery inverters and STATCOMs has been shown to raise data-center hosting capacity, but the study was limited to a distribution feeder~\cite{fang2026hosting}.

This paper presents a multi-period AC OPF that couples a transmission network with a physically based AIDC model driven by measured GPU power traces~\cite{vercellino2026nlr} to evaluate the grid impact of the facility and the value of its flexibility. The model represents deadline-constrained workload classes, the thermal mass of the hall, a BESS with reactive capability, behind-the-meter PV, and an optional STATCOM. Using the RTS-GMLC test system~\cite{barrows2020rts} at five-minute resolution, we examine how the AIDC affects congestion, bus voltages, and operating cost across utilization levels, how far workload, cooling, and battery flexibility mitigate these effects, and whether reactive support from the battery inverter or a STATCOM restores the POI voltage. The results separate active-power flexibility, which sets the cost and congestion impact, from reactive-power flexibility, which sets the POI voltage, and show how workload shifting and BESS sizing interact.

\section{System Model and Problem Formulation}
\label{sec:model}

In the presented model, the host grid is operated over a horizon $\mathcal{T}=\{1,\dots,T\}$ of five-minute periods, the native resolution of the test-system data~\cite{barrows2020rts}, by a multi-period AC-OPF. Sets are buses $\mathcal{N}$, branches $\mathcal{L}$ (branch $l$ from bus $i$ to bus $j$), and generators $\mathcal{G}$, split into thermal units $\mathcal{G}^{\mathrm{th}}$, curtailable renewables $\mathcal{G}^{\mathrm{cu}}$ (utility PV, wind, CSP), fixed-profile units $\mathcal{G}^{\mathrm{fx}}$ (hydro, rooftop PV), and synchronous condensers; $\mathcal{G}_i$ are the units at bus~$i$.

\subsection{AI Data Center Model}
\label{sec:aidc}
The AIDC at bus $a$ comprises IT equipment, a cooling plant, a BESS, and PV behind the POI.

The facility hosts inference and training in separate server pools $w\in\{\mathrm{inf},\mathrm{tr}\}$ holding shares $\alpha_w$ of the rated IT power $S^{\mathrm{IT}}$, which is set by the 3:2 inference-to-training energy split as reported for a hyperscale machine-learning fleet~\cite{patterson2022carbon}. Each pool follows one-minute traces of an LLM-inference facility or of a training and fine-tuning facility, simulated from measured GPU power profiles at a given average utilization level~\cite{vercellino2026nlr}. Averaged over each period, the traces give the IT power $\pi^{w}_t$ as a share of the rated IT design and the utilization $u^{0}_{w,t}$, the fraction of servers busy.
Over a year of data, IT power is affine in utilization with slope $\beta^{w}$ ($R^2 \ge 0.999$). The served utilization $u_{w,t}$ may depart from $u^{0}_{w,t}$ within the bounds (\ref{eq:band}), and the IT power follows (\ref{eq:it}).
\begin{equation}
P^{\mathrm{it}}_t = \sum_{w} \alpha_w S^{\mathrm{IT}}\big[\pi^{w}_t + \beta^{w}(u_{w,t} - u^{0}_{w,t})\big]. \label{eq:it}
\end{equation}
\begin{equation}
0 \le u_{w,t} \le 1, \qquad |u_{w,t} - u^{0}_{w,t}| \le \rho, \label{eq:band}
\end{equation}
where the shift band $\rho$ limits how far the served work may depart from the arriving work in each period.

 The arriving work is split into classes $k\in\mathcal{K}$ in which inference is served by the inference pool, and short and long training are served in the training pool. Class $k$ receives the share $\chi_k$ of the arrivals $u^{0}_{w(k),t}$ of its pool $w(k)$, has deadline $D_k$ in periods, is served at rate $s_{k,t}$, may be shed at rate $\sigma_{k,t}$, and accumulates backlog $B_{k,t}$ according to (\ref{eq:backlog}).
\begin{subequations}\label{eq:classes}
\begin{align}
B_{k,t+1} &= B_{k,t} + \chi_k u^{0}_{w(k),t} - s_{k,t} - \sigma_{k,t}, \label{eq:backlog}\\
B_{k,t+1} &\le \textstyle\sum_{\tau=\max(1,\,t-D_k+1)}^{t} \chi_k u^{0}_{w(k),\tau}, \label{eq:deadline}\\
u_{w,t} &= \textstyle\sum_{k:\,w(k)=w} s_{k,t}, \qquad B_{k,1} = B_{k,T+1} = 0, \label{eq:classsum}\\
& B_{k,t},\ s_{k,t},\ \sigma_{k,t} \ge 0. \label{eq:classpos}
\end{align}
\end{subequations}
The deadline constraint \eqref{eq:deadline} keeps the backlog below the work that arrived within the last $D_k$ periods. Work within a class is served first-in, first-out, so \eqref{eq:deadline} holds exactly and deferred work is served within its deadline, and (\ref{eq:classsum}) ties the served classes to their pool and clears the backlog at both ends of the day. Inference has $D_k = 0$ and is served on arrival or shed.

 The hall is a lumped thermal mass $C^{\mathrm{th}}$ heated by the IT load, whose electrical power is converted entirely into heat ($1~\mathrm{MW} = 1~\mathrm{MW_{th}}$), and cooled by heat extraction $Q_t$, whose electrical draw is set by an energy-input ratio $\epsilon$, as modeled in (\ref{eq:cooling}).
\begin{subequations}\label{eq:cooling}
\begin{align}
T^{\mathrm{in}}_{t+1} &= T^{\mathrm{in}}_t + \tfrac{\Delta t}{C^{\mathrm{th}}}\big(P^{\mathrm{it}}_t - Q_t\big), \qquad P^{\mathrm{cool}}_t = \epsilon\, Q_t, \label{eq:hall}\\
0 \le Q_t &\le \overline{Q}, \quad \underline{T}^{\mathrm{in}} \le T^{\mathrm{in}}_t \le \overline{T}^{\mathrm{in}}, \quad T^{\mathrm{in}}_1 = T^{\mathrm{in}}_{T+1} = T^{\mathrm{in},0}, \label{eq:hallbounds}
\end{align}
\end{subequations}
where the end condition prevents deferring cooling energy beyond the horizon, and the hall is treated as adiabatic over the day.
The charge and discharge of the BESS are limited by (\ref{eq:bessp}), and the state of energy, with charging and discharging efficiencies $\eta^{\mathrm{ch}}, \eta^{\mathrm{dis}}$, is defined in (\ref{eq:soc}). The BESS state of energy and the onsite PV output, up to its availability $P^{\mathrm{pv,av}}_t$, are bounded in (\ref{eq:ebounds}).
\begin{subequations}\label{eq:bess}
\begin{align}
&0 \le P^{\mathrm{ch}}_t \le \overline{P}^{\mathrm{b}}, \qquad 0 \le P^{\mathrm{dis}}_t \le \overline{P}^{\mathrm{b}}, \label{eq:bessp}\\
E_{t+1} &= E_t + \Delta t\big(\eta^{\mathrm{ch}} P^{\mathrm{ch}}_t - P^{\mathrm{dis}}_t/\eta^{\mathrm{dis}}\big), \quad E_{T+1} = E_1, \label{eq:soc}\\
\underline{E} &\le E_t \le \overline{E}, \qquad 0 \le P^{\mathrm{pv}}_t \le P^{\mathrm{pv,av}}_t, \label{eq:ebounds}\\
&|q^{\mathrm{b}}_t| \le \tan\!\big(\arccos 0.95\big)\,\overline{P}^{\mathrm{b}}. \label{eq:bessq}
\end{align}
\end{subequations}
Constraint \eqref{eq:bessq} lets the BESS inverter inject or absorb reactive power $q^{\mathrm{b}}_t$ over the $\pm 0.95$ power-factor range at rated power that FERC Order No.~827 requires of non-synchronous generators~\cite{ferc2016order827}.
 The inverter is rated $\overline{P}^{\mathrm{b}}/0.95$, so its apparent-power limit never binds, and the PV runs at unity power factor to isolate the voltage effect of the BESS and the STATCOM.

 A STATCOM of rating $\overline{Q}^{\mathrm{st}}$ injects reactive power $q^{\mathrm{st}}_t$ and is modeled with negligible losses. The AIDC draw and its ramp limit $\overline{R}^{\mathrm{poi}}$ per five-minute period, as recommended for large loads~\cite{nerc2026rglargeloads}, are given in (\ref{eq:poi}).
\begin{subequations}\label{eq:poi}
\begin{align}
P^{\mathrm{dc}}_t &= P^{\mathrm{it}}_t + P^{\mathrm{cool}}_t + P^{\mathrm{ch}}_t - P^{\mathrm{dis}}_t - P^{\mathrm{pv}}_t, \label{eq:aidc}\\
Q^{\mathrm{dc}}_t &= \big(P^{\mathrm{it}}_t + P^{\mathrm{cool}}_t\big)\tan\varphi - q^{\mathrm{b}}_t - q^{\mathrm{st}}_t, \quad |q^{\mathrm{st}}_t| \le \overline{Q}^{\mathrm{st}}, \label{eq:aidcq}\\
\big|P^{\mathrm{dc}}_t &- P^{\mathrm{dc}}_{t-1}\big| \le \overline{R}^{\mathrm{poi}}, \label{eq:poiramp}
\end{align}
\end{subequations}
with lagging power factor $\cos\varphi$~\cite{gridlab2025largeloads}. The inflexible AIDC keeps its PV but has no other flexibility, as fixed by (\ref{eq:inflex}).
\begin{subequations}\label{eq:inflex}
\begin{align}
&u_{w,t} = u^{0}_{w,t}, \quad \sigma_{k,t} = B_{k,t} = 0, \quad Q_t = P^{\mathrm{it}}_t, \label{eq:inflexfix}\\
&P^{\mathrm{ch}}_t = P^{\mathrm{dis}}_t = q^{\mathrm{b}}_t = q^{\mathrm{st}}_t = 0, \label{eq:inflexbess}\\
&P^{\mathrm{it}}_t + P^{\mathrm{cool}}_t = (1+\epsilon)\, S^{\mathrm{IT}} \textstyle\sum_w \alpha_w \pi^{w}_t. \label{eq:inflexload}
\end{align}
\end{subequations}
With $\epsilon = \mathrm{PUE}-1$, constraint \eqref{eq:inflexload} is the facility power implied by the power usage effectiveness (PUE)~\cite{shehabi2024lbnl}. 

In this paper, a flexible AIDC is one that can reshape its net draw at the POI while still serving all its computing work within its deadlines. It does so by deferring training work, pre-cooling the hall, and cycling the BESS.

\subsection{Multi-Period AC Optimal Power Flow}
\label{sec:acopf}
For every $t\in\mathcal{T}$, the decision variables are the bus voltages $V_{i,t} = v_{i,t}e^{\mathrm{j}\theta_{i,t}}$, the generator outputs $S_{g,t} = p_{g,t} + \mathrm{j}q_{g,t}$, the branch flows $S^{\mathrm{fr}}_{l,t}, S^{\mathrm{to}}_{l,t}$ at both ends, the HVDC flows, and the reserves $r_{g,h,t}$. With the $\Pi$-model admittances $Y^l_{ii}, Y^l_{ij}, Y^l_{ji}, Y^l_{jj}$ of branch $l$ from bus $i$ to bus $j$ (series admittance, line charging, and off-nominal tap), the AC power flow (\ref{eq:flows})-(\ref{eq:flowsto}) and the nodal balance (\ref{eq:balance}) are

\begin{subequations}\label{eq:acpf}
\begin{align}
S^{\mathrm{fr}}_{l,t} &= (Y^l_{ii})^{*} v_{i,t}^2 + (Y^l_{ij})^{*} V_{i,t} V_{j,t}^{*}, \label{eq:flows}\\
S^{\mathrm{to}}_{l,t} &= (Y^l_{jj})^{*} v_{j,t}^2 + (Y^l_{ji})^{*} V_{j,t} V_{i,t}^{*}, \label{eq:flowsto}\\
\textstyle\sum_{g\in\mathcal{G}_i} S_{g,t} - S^{\mathrm{out}}_{i,t} &= S^{\mathrm{D}}_{i,t} + (Y^{\mathrm{sh}}_i)^{*} v_{i,t}^2 + \mathbf{1}_{i=a}
S^{\mathrm{dc}}_t, \label{eq:balance}
\end{align}
\end{subequations}
where $S^{\mathrm{out}}_{i,t}$ sums the branch flows at their ends at bus $i$ and the HVDC injection there, $S^{\mathrm{D}}_{i,t}$ is the demand, $Y^{\mathrm{sh}}_i$ the shunt admittance, $S^{\mathrm{dc}}_t = P^{\mathrm{dc}}_t + \mathrm{j}Q^{\mathrm{dc}}_t$ the AIDC draw, and $\mathbf{1}_{i=a}$ equals one at the AIDC bus $a$ and zero otherwise. The dual of the active part of (\ref{eq:balance}) is the locational marginal price (LMP). The network and generator limits are
\begin{subequations}\label{eq:netlim}
\begin{align}
&\theta_{\mathrm{ref},t} = 0, \quad \underline{V}_i \le v_{i,t} \le \overline{V}_i, \quad \underline{\theta}_l \le \theta_{i,t} - \theta_{j,t} \le \overline{\theta}_l, \label{eq:vlim}\\
&|\mathrm{Re}\,S^{\mathrm{fr}}_{l,t}|, |\mathrm{Re}\,S^{\mathrm{to}}_{l,t}| \le \overline{P}_l, \quad \underline{Q}_g \le q_{g,t} \le \overline{Q}_g, \label{eq:linelim}\\
&0 \le p_{g,t} \le \overline{P}_{g,t}, \quad |p_{g,t} - p_{g,t-1}| \le R_g \ \ (g\in\mathcal{G}^{\mathrm{th}}), \label{eq:genlim}
\end{align}
\end{subequations}
where $\overline{P}_{g,t}$ is the capacity $\overline{P}_g$ of a thermal unit or the available output $P^{\mathrm{av}}_{g,t}$ of a curtailable renewable, fixed-profile units produce $P^{\mathrm{av}}_{g,t}$, and $R_g$ is the ramp limit per period (MW/5~min). The branch limit is on active power because the test system rates its branches as MW flow limits~\cite{barrows2020rts}, and the HVDC flows follow the linear loss model of the link within its rating. Each reserve product $h\in\mathcal{H}$ (regional spinning, regulation up and down~\cite{barrows2020rts}) with requirement $D^{\mathrm{r}}_{h,t}$, deployment window $\tau_h$ in periods, and eligible set $\mathcal{G}_h$ satisfies
\begin{subequations}\label{eq:reserve}
\begin{align}
&\textstyle\sum_{g\in\mathcal{G}_h} r_{g,h,t} \ge D^{\mathrm{r}}_{h,t}, \qquad 0 \le r_{g,h,t} \le R_g \tau_h \ (g\in\mathcal{G}^{\mathrm{th}}), \label{eq:resreq}\\
&p_{g,t} + \textstyle\sum_{h\in\mathcal{H}^{\uparrow}} r_{g,h,t} \le \overline{P}_{g,t}, \qquad p_{g,t} - \textstyle\sum_{h\in\mathcal{H}^{\downarrow}} r_{g,h,t} \ge 0, \label{eq:reshead}
\end{align}
\end{subequations}
where $\mathcal{H}^{\uparrow}$ and $\mathcal{H}^{\downarrow}$ are the upward and downward products. The AIDC is not enrolled as a reserve provider, so the reserves come from the grid alone.

The objective (\ref{eq:obj}) is that of a central operator that co-optimizes the grid and the AIDC. It minimizes the production cost of the grid, the AIDC costs of shedding computing work and of BESS degradation, and a penalty on changes in the AIDC draw.
\begin{multline}
\min \ \sum_{t\in\mathcal{T}} \Delta t \Big[\!\sum_{g\in\mathcal{G}^{\mathrm{th}}} c_{g,t}\, p_{g,t} + c^{\mathrm{shed}} S^{\mathrm{IT}}\sum_{k}\alpha_{w(k)}\beta^{w(k)}\sigma_{k,t} \\
+ c^{\mathrm{deg}}\big(\eta^{\mathrm{ch}}P^{\mathrm{ch}}_t + P^{\mathrm{dis}}_t/\eta^{\mathrm{dis}}\big) + \gamma\big(P^{\mathrm{dc}}_t - P^{\mathrm{dc}}_{t-1}\big)^2\Big], \label{eq:obj}
\end{multline}
where $\Delta t = 1/12$~h is the period length, $c_{g,t} = m_t c_g$ scales the marginal cost $c_g$ of unit $g$ by a time-varying price shape $m_t$ with a daily mean of one, $c^{\mathrm{shed}}$ and $c^{\mathrm{deg}}$ are the shed penalty and the BESS degradation cost, and $\gamma$ weights the change in AIDC draw between periods. Whenever the AIDC has a dispatchable resource, the $\gamma$ term selects, among schedules of nearly equal cost, the one with the smoothest POI draw.

\begin{figure}[!t]
\centering
\includegraphics[width=0.8\columnwidth]{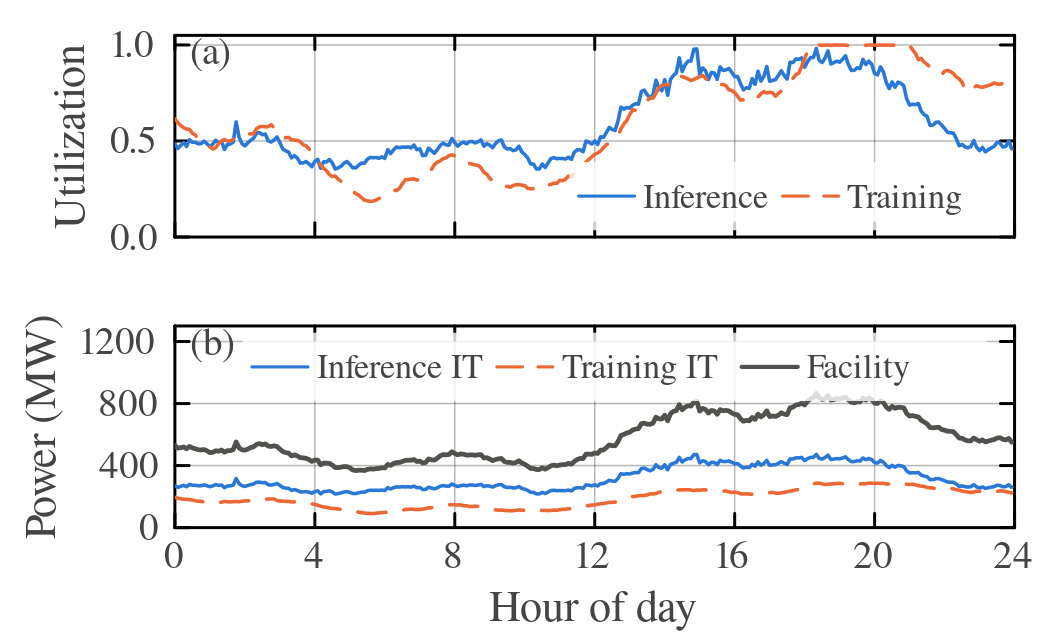}
\caption{AIDC at the nominal 60\% utilization. (a) Utilization of the inference and training pools. (b) IT power of each pool and facility load before PV.}
\label{fig:profiles}
\end{figure}

\section{Case Study and Results}
\label{sec:case}
The presented setup is simulated on the RTS-GMLC test system~\cite{barrows2020rts}, with 73 buses, 120 branches, an HVDC link, and 158 units, including utility and rooftop PV, wind, and hydro, and with its published five-minute real-time load and renewable profiles. The selected data is drawn from 26 August 2020, the peak real-time load day of the data set (7,977~MW). The price shape $m_t$ is the hourly CAISO SP15 day-ahead price of 26 August 2026~\cite{caiso2026sp15}.
The model is implemented in JuMP and solved with Ipopt at a tolerance of $10^{-6}$ for convergence, and each simulation run of 288 periods takes 2 to 10~min. The AIDC is connected at bus~119. Its rated IT power is $S^{\mathrm{IT}} = 1{,}000$~MW, split 0.6/0.4 between the inference and training pools~\cite{patterson2022carbon}, with slopes $\beta^{\mathrm{inf}} = 0.679$ and $\beta^{\mathrm{tr}} = 0.586$ fitted to~\cite{vercellino2026nlr}, a PUE of 1.14 ($\epsilon = 0.14$)~\cite{shehabi2024lbnl}, and a 0.95 lagging power factor~\cite{gridlab2025largeloads}. Inference, short training, and long training receive 60\%, 12\%, and 28\% of the work, with deadlines of 0, 60, and 240~min, a shift band $\rho = 0.15$, and a shed penalty of \$10{,}000/MWh. The hall has $C^{\mathrm{th}} = 240$~MWh$_{\mathrm{th}}$/$^{\circ}$C and $\overline{Q} = 1{,}000$~MW$_{\mathrm{th}}$, which are half the per-MW capacitance and the same per-MW heat extraction of~\cite{lu2026battery}, and is held between 18 and 27~$^{\circ}$C~\cite{ashrae2021thermal}, starting at 22~$^{\circ}$C. The BESS is rated 500~MW and 1{,}000~MWh, with a 283~MWh floor, which represents 20\% of the rated energy capacity plus five minutes of rated IT power kept as an uninterruptible ride-through reserve. Charge and discharge efficiencies are 0.95, reactive limit is $\pm$164~MVAr, and a degradation cost of \$9 per MWh of one-way throughput, applied to both charging and discharging, which is half of the \$18/MWh per full cycle implied by a \$108/kWh pack price~\cite{bnef2025battery} over 6{,}000 cycles. The PV is rated 300~MW, the STATCOM at 300~MVAr, and the POI ramp limit is set to 150~MW/5-min (30~MW/min), with $\gamma = 0.1$~\$/h per MW$^2$. The nominal case uses the traces of~\cite{vercellino2026nlr} at 60\% average utilization. As Fig.~\ref{fig:profiles}(a) shows, the inference pool follows a diurnal request pattern, while the training pool switches between job-driven plateaus. Before PV, the facility load in Fig.~\ref{fig:profiles}(b) ranges from 368 to 865~MW. The 300~MW PV follows the profile of the RTS-GMLC PV plant at the same bus. The 20\%, 40\%, and 80\% profiles serve as a sensitivity on utilization.

\subsection{AIDC Operation at Nominal Utilization}
\label{sec:results}
\begin{figure}[!t]
\centering
\includegraphics[width=\columnwidth]{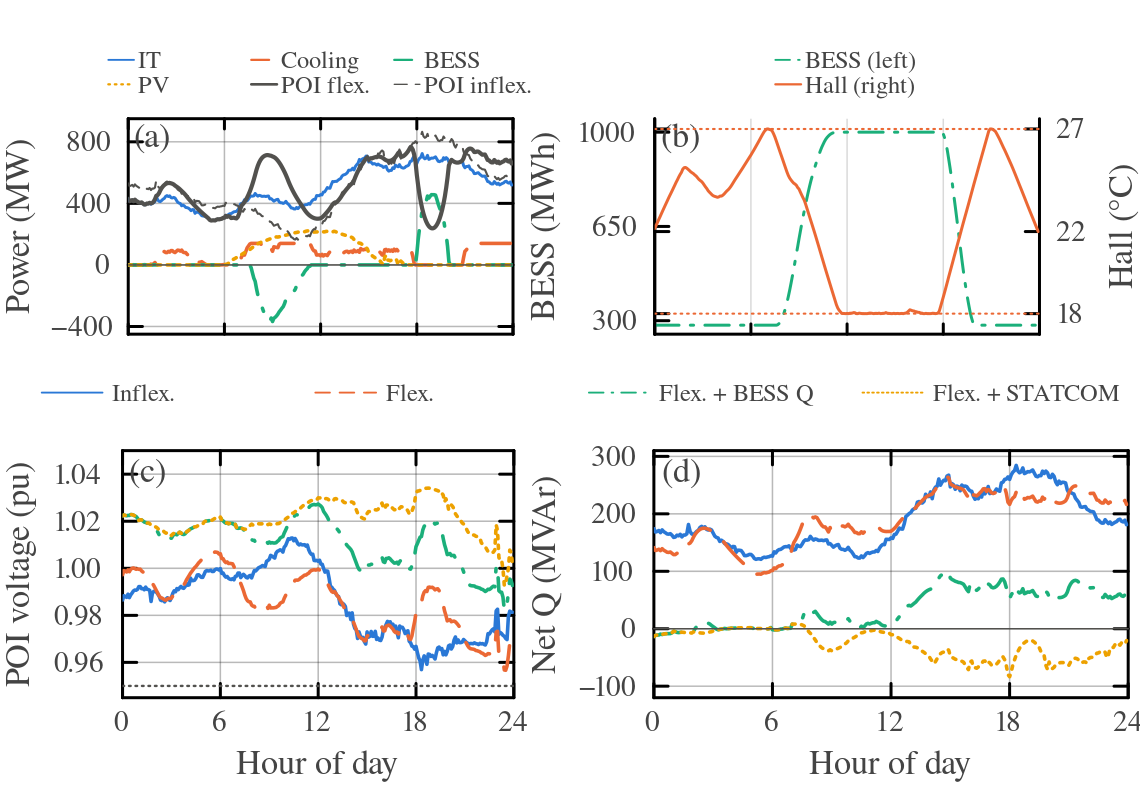}
\caption{AIDC at 60\% utilization. (a) Flexible AIDC dispatch and POI draw. (b) BESS energy and hall temperature. (c) Voltage and (d) net reactive power drawn from the grid at the POI (bus 119). The dotted line in (c) is the 0.95~pu limit.}

\label{fig:dispatch}
\end{figure}

The presented flexibility lowers the cost, the congestion, and the price spike that the AIDC causes at its POI, while reactive support restores the POI voltage. Fig.~\ref{fig:dispatch} compares four cases at 60\% utilization. The \textbf{inflexible} AIDC serves the IT workload as it arrives and can only curtail its PV. The \textbf{flexible} AIDC shifts training work, precools the hall, and cycles the BESS, with $q^{\mathrm{b}}_t = q^{\mathrm{st}}_t = 0$. The \textbf{flexible + BESS Q} case adds the BESS reactive power within (\ref{eq:bessq}), and the \textbf{flexible + STATCOM} case adds the 300~MVAr STATCOM instead, with $q^{\mathrm{b}}_t = 0$. The inflexible AIDC raises the daily operating cost from \$2.147M without the AIDC to \$2.531M, or \$31.3 per MWh drawn by the facility. It draws up to 284~MVAr in the evening peak, which lowers the POI voltage to 0.957~pu, and holds branch A28 next to the POI at its limit in 96 periods. We report a branch as binding, a stress indicator rather than a violation, when its loading reaches 99\% of its rating.
 Compared with the inflexible AIDC, the flexible AIDC lowers daily cost by \$43.8k (a 1.73\% drop) and reduces A28 binding periods from 96 to 90. As Fig.~\ref{fig:dispatch}(a) shows, it raises its draw to about 710~MW in the cheapest hours after 08:00 to serve deferred training, precool the hall, and charge the BESS, which is full by 11:20. The hall is then held at 18~$^{\circ}$C until 17:40 in Fig.~\ref{fig:dispatch}(b). From 18:00 to 20:05, across the day-ahead price peak, the BESS discharges 681~MWh and cooling stops while the hall warms toward 27~$^{\circ}$C. Training is deferred from the night to the morning and from the evening peak to the last hours of the day, and the mean evening draw falls by 20.8\%. This shift also removes most of the price spike the facility creates at its bus. The peak LMP at bus~119 rises from \$59.1/MWh without the AIDC to \$87.1/MWh with the inflexible AIDC and falls to \$62.5/MWh with flexibility, while the daily mean changes by less than \$0.3/MWh.

As Fig.~\ref{fig:dispatch}(c) shows, flexibility alone makes the POI voltage more variable but leaves its minimum at 0.957~pu, well below nominal and close to the 0.95~pu limit. The dispatch minimizes cost and treats voltage only as a limit, so it lowers the voltage by up to 0.023~pu during morning charging and raises it by up to 0.031~pu in the evening. The BESS inverter, operated at $\pm$0.95 power factor, supplies up to 164~MVAr locally, which cuts the reactive draw from the grid in Fig.~\ref{fig:dispatch}(d) and raises the minimum POI voltage to 0.982~pu and its daily mean from 0.985 to 1.011~pu. The STATCOM gives the best profile, with a minimum of 0.993~pu, because it makes the facility a net reactive source in 86\% of the periods. Its net reactive draw from the grid never exceeds 9~MVAr, against up to 284~MVAr for the inflexible AIDC, so the grid no longer supplies the facility's reactive power and the margin above the 0.95~pu limit grows from 0.007 to 0.043~pu. Both lower the daily cost of the flexible AIDC slightly, by \$1.0k and \$1.2k.

\subsection{Impact of Utilization Level}
The grid impact of the AIDC grows with the utilization, and the proposed flexibility model relieves congestion wherever it appears. Table~\ref{tab:util} compares the inflexible AIDC with the flexible AIDC with BESS reactive support at four utilization levels, with no limit violated in any case. A28 binds from 40\% upward and for half of the day at 80\%, where A11 also binds and the POI voltage approaches 0.95~pu. Flexibility lowers the cost by \$38--45k per day and raises the minimum POI voltage by 0.014--0.026~pu at every level. Moreover, it cuts the binding periods of A28 from 18 to 6 at 40\%, from 96 to 91 at 60\%, and from 143 to 135 at 80\%, where those of A11 also fall from 23 to 14. At 20\%, flexibility raises the peak POI draw from 344 to 536~MW to charge the BESS and precool the hall, yet branch A28 stays below its limit. At 80\%, flexibility lowers the peak LMP at bus~119 from \$80.1 to \$62.4/MWh. Notably, this inflexible peak is below the \$87.1/MWh at 60\%, although the mean LMP is higher at 80\%, both from 18:00 to 20:00 (\$73.0 vs.\ \$67.3/MWh) and over the day (\$30.4 vs.\ \$28.8/MWh). The 60\% draw steps up by 27~MW to its daily maximum at 18:20, whereas the 80\% draw stays near 870~MW, so the spike follows the step in the draw rather than its level.

\begin{table}[!t]
\centering
\caption{Inflexible / Flexible AIDC Across Utilization Levels}
\label{tab:util}
\footnotesize
\setlength{\tabcolsep}{3.5pt}
\begin{tabular}{@{}c c c c c@{}}
\hline
Utilization & Cost (\$M/day) & A28 binding & A11 binding & Min. POI $V$ (pu) \\
\hline
20\% & 2.305 / 2.267 & 0 / 0 & 0 / 0 & 0.998 / 1.011 \\
40\% & 2.420 / 2.377 & 18 / 6 & 0 / 0 & 0.981 / 1.001 \\
60\% & 2.531 / 2.487 & 96 / 91 & 1 / 0 & 0.957 / 0.982 \\
80\% & 2.620 / 2.578 & 143 / 135 & 23 / 14 & 0.953 / 0.980 \\
\hline
\end{tabular}\\[2pt]
\begin{minipage}{\columnwidth}\scriptsize
Binding: periods (of 288) with branch loading $\ge$ 99\% of its rating. Flexible: workload, cooling, BESS with reactive support, and PV.
\end{minipage}
\end{table}

\begin{table}[!t]
\centering
\caption{BESS Size at 60\% and 80\% Utilization}
\label{tab:bess}
\footnotesize
\setlength{\tabcolsep}{3pt}
\begin{tabular}{@{}c c c c c c c c c@{}}
\hline
 & \multicolumn{4}{c}{60\%} & \multicolumn{4}{c}{80\%} \\
BESS & Saving & A28 & $V_{\min}$ & $P^{\mathrm{dc}}_{\max}$ & Saving & A28 & $V_{\min}$ & $P^{\mathrm{dc}}_{\max}$ \\
(MW $\times$ h) & (\$k) & & (pu) & (MW) & (\$k) & & (pu) & (MW) \\
\hline
None & -- & 110 & 0.957 & 766 & -- & 129 & 0.954 & 870 \\
125 $\times$ 2 & 5.4 & 94 & 0.964 & 765 & 5.5 & 129 & 0.962 & 870 \\
125 $\times$ 4 & 12.7 & 90 & 0.964 & 768 & 12.9 & 110 & 0.962 & 871 \\
250 $\times$ 2 & 13.9 & 93 & 0.970 & 765 & 14.1 & 111 & 0.969 & 870 \\
250 $\times$ 4 & 27.0 & 84 & 0.970 & 760 & 27.2 & 133 & 0.969 & 869 \\
500 $\times$ 1 & 14.3 & 93 & 0.982 & 765 & 14.7 & 112 & 0.980 & 870 \\
500 $\times$ 2 & 30.2 & 91 & 0.982 & 763 & 30.3 & 135 & 0.980 & 870 \\
500 $\times$ 4 & 53.6 & 97 & 0.982 & 829 & 53.0 & 162 & 0.980 & 942 \\
750 $\times$ 2 & 45.6 & 97 & 0.993 & 788 & 45.3 & 154 & 0.990 & 916 \\
750 $\times$ 4 & 78.2 & 120 & 0.993 & 926 & 76.2 & 161 & 0.990 & 1019 \\
\hline
\end{tabular}\\[2pt]
\begin{minipage}{\columnwidth}\scriptsize
Saving: daily cost reduction relative to the same flexible AIDC without a BESS. A28: binding periods (of 288). $V_{\min}$: minimum POI voltage. $P^{\mathrm{dc}}_{\max}$: peak POI draw. Flexible AIDC with BESS reactive support ($\pm$0.95 pf), $\rho = 0.15$, and without STATCOM.
\end{minipage}
\end{table}

\subsection{Impact of BESS Size}
The BESS energy sets the cost saving and its power rating the voltage benefit, while a cost-driven BESS relieves congestion only when the network has spare capacity during charging. Table~\ref{tab:bess} varies the BESS power and duration of the flexible AIDC with BESS reactive support and reports the saving relative to the same AIDC without a BESS. At 60\%, workload and cooling flexibility alone save \$14.7k per day relative to the inflexible AIDC, and the 500~MW $\times$ 2~h BESS with reactive support adds \$30.2k, for a total of \$44.8k.
The cost saving grows almost in proportion to the stored energy, from \$5.4k per day for 125~MW $\times$ 2~h to \$78.2k for 750~MW $\times$ 4~h at 60\%, and is nearly the same at 80\%. The voltage benefit depends only on the power rating, since the reactive limit scales with it, and 500~MW raises the minimum POI voltage to 0.982~pu at 60\% and 0.980~pu at 80\% for any duration. Without a BESS, workload and cooling flexibility raise the binding periods of A28 at 60\% from 96 to 110, because deferred training is served in the late evening while A28 is still loaded. Every BESS of up to 2,000~MWh brings them back to 84--97, but the 750~MW $\times$ 4~h unit charges in the morning at up to 926~MW and raises them to 120. At 80\%, A28 is near its rating for most of the day. Units of up to 500~MWh keep its binding periods at or below the 129 of the no-BESS case, whereas units of 1,000~MWh or more add up to 33. Furthermore, the largest units also raise the peak POI draw, by up to 21.0\% at 60\% and 17.1\% at 80\%.

\begin{figure}[!t]
\centering
\includegraphics[width=0.96\columnwidth]{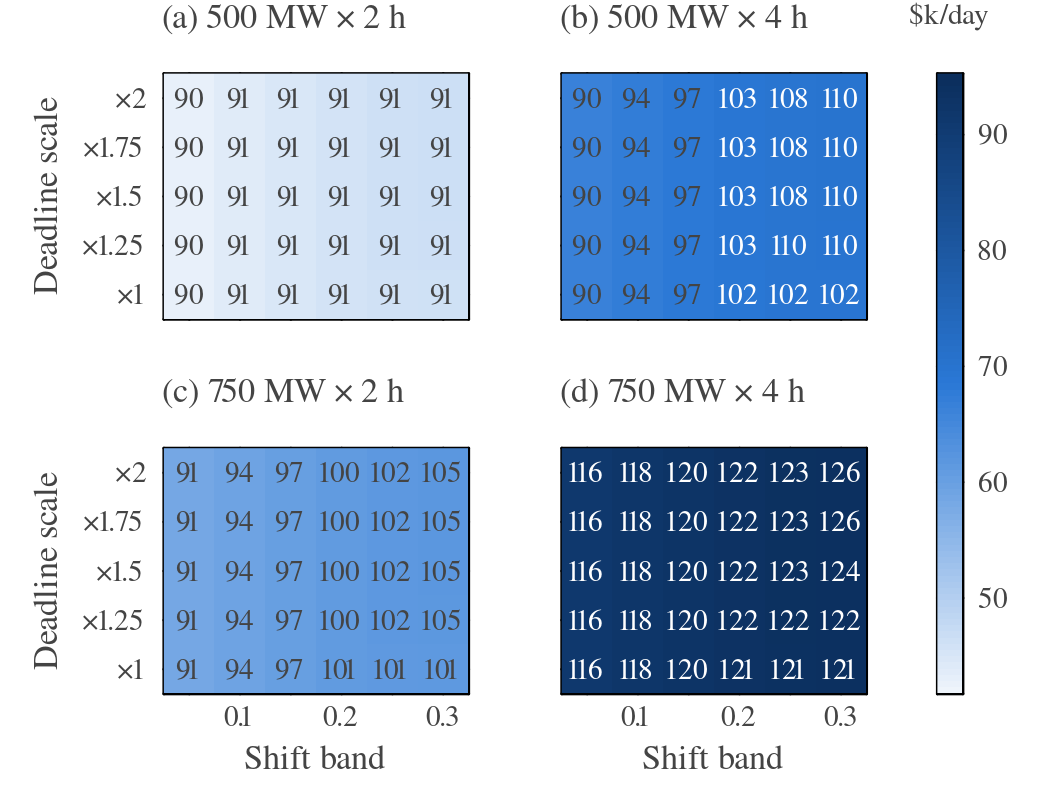}
\caption{Shift band $\rho$ and deadline scale of the training classes at 60\% utilization for four BESS sizes, flexible AIDC with BESS reactive support. Color: daily savings relative to the inflexible AIDC. Numbers: A28 binding periods.}
\label{fig:grid}
\end{figure}

\subsection{Impact of Shift Band and Deadlines}
Increasing the shift band adds a modest saving with diminishing returns; the deadlines barely matter, and a larger BESS leaves less value for workload shifting. Fig.~\ref{fig:grid} varies the shift band $\rho$ in (\ref{eq:band}) from 0.05 to 0.30 and scales the deadlines of both training classes by 1.25 to 2 for four BESS sizes, with the saving measured against the inflexible AIDC. The saving rises at every step of $\rho$, because a wider band lets more training work leave the expensive evening hours, but each step adds less than the one before, from \$1.0k for the first step to \$0.1k for the last with the 500~MW $\times$ 2~h BESS in Fig.~\ref{fig:grid}(a). For $\rho \le 0.15$, increasing the deadlines leaves the saving unchanged for every BESS size, and above it, doubling them adds at most \$0.4k per day. The shift band, rather than the deadlines, therefore limits how much work is deferred.

The BESS sets the level of the saving. At the nominal $\rho = 0.15$, the saving is \$44.8k per day with 500~MW $\times$ 2~h (1,000~MWh) and \$60.3k with 750~MW $\times$ 2~h in Fig.~\ref{fig:grid}(c). With the four-hour units, it reaches \$68.3k with 500~MW $\times$ 4~h in Fig.~\ref{fig:grid}(b) and \$92.9k with 750~MW $\times$ 4~h (3,000~MWh) in Fig.~\ref{fig:grid}(d). Increasing the band from 0.05 to 0.30 adds \$3.4--3.7k per day with the three smaller units and \$2.8k with 750~MW $\times$ 4~h, so the value of workload shifting shrinks as the BESS grows. The wider band also adds congestion with every BESS larger than 500~MW $\times$ 2~h. With that unit, A28 binds in 90--91 periods at every setting, whereas its binding periods rise with $\rho$ from 91 to 105 with 750~MW $\times$ 2~h, from 90 to 110 with 500~MW $\times$ 4~h, and from 116 to 126 with 750~MW $\times$ 4~h.  Most of the added periods fall between 06:00 and 12:00, when the BESS charges and the deferred training is being served.

\section{Conclusion}
This paper studies the grid impact of a gigawatt-scale flexible AIDC with a multi-period AC OPF on the RTS-GMLC system. The case studies compare inflexible and flexible operation across utilization levels, BESS sizes, shift bands, and workload deadlines, with reactive support from the BESS or a STATCOM. An inflexible AIDC raises the operating cost and pulls the POI voltage toward 0.95~pu. From 40\% utilization upward, it also binds the line next to the POI, for half of the day at 80\%. Flexibility with BESS reactive support lowers the daily cost by \$38--45k and relieves this congestion. Without reactive support, workload, cooling, and battery flexibility leave the minimum POI voltage unchanged. Reactive support raises it to 0.982~pu with the BESS and 0.993~pu with a STATCOM. The BESS energy sets the saving, and its power rating sets the voltage benefit. When the BESS is dispatched for cost, the largest capacity binds the line more often than the inflexible AIDC. With a BESS on site, widening the shift band yields diminishing returns. As a stress case, the study bounds the impacts a host grid would see at such an interconnection.

\end{document}